\documentclass[twocolumn, amsmath,amssymb,aps, prbl,
showkeys,showpacs, floatfix, superscriptaddress] {revtex4-2}
\usepackage{times}
\usepackage{graphicx}
\usepackage[usenames,dvipsnames]{color}
\usepackage{bm}
\usepackage{amsmath}
\usepackage{amssymb}

\usepackage[colorlinks=true,breaklinks=true,allcolors=blue]{hyperref}

\begin{document}

\title{Boundary Conditions for the Entanglement Cut in 2D Conformal Field Theories} 
\author{Ananda Roy}
\email{ananda.roy@physics.rutgers.edu}
\affiliation{Department of Physics and Astronomy, Rutgers University, Piscataway, NJ 08854-8019 USA}
\author{Sergei L. Lukyanov}
\affiliation{Department of Physics and Astronomy, Rutgers University, Piscataway, NJ 08854-8019 USA}
\author{Hubert Saleur}
\affiliation{Institut de physique théorique, CEA, CNRS, Université Paris-Saclay, France}
\affiliation{Physics Department, University of Southern California, Los Angeles, USA}

\begin{abstract}
The entanglement spectra for a subsystem in a spin chain fine-tuned to a quantum-critical point contains signatures of the underlying quantum field theory that governs its low-energy properties. For an open chain with given boundary conditions  described by a 2D conformal field theory~(CFT), the entanglement spectrum of the left/right half of the system coincides with a boundary CFT spectrum, where one of the boundary conditions arise due to the `entanglement cut'. The latter has been argued to be conformal and has been numerically found to be the `free' boundary condition for Ising, Potts and free boson theories. For these models, the `free' boundary condition for the lattice degree of freedom has a counterpart in the continuum theory. However, this is not true in general. Here, this question is analyzed for the unitary minimal models of 2D CFTs using the density matrix renormalization group technique. The entanglement spectra are computed for blocks of  spins in open chains of A-type restricted solid-on-solid models with identical boundary conditions at the ends. The imposed boundary conditions are realized exactly for these lattice models due to their  integrable nature. The obtained entanglement spectra are in good agreement with certain boundary CFT spectra. The boundary condition for the entanglement cut is found to be conformal and to coincide with the one with the highest boundary entropy. This identification enables determination of the exponents governing the unusual corrections to the entanglement entropy from the CFT partition functions. These are compared with numerical results. 
\end{abstract}

\maketitle 
The entanglement Hamiltonian is important for the analysis of correlations in quantum many body systems and quantum field theories~\cite{bisognano1975duality, bisognano1976duality, Peschel1999, Li2008}. For a subsystem~$A$ with reduced density matrix~$\rho_A$, the entanglement Hamiltonian is defined as
\begin{equation}
K_A = -\frac{1}{2\pi}\ln\rho_A.
\end{equation}
For ground states of one-dimensional quantum critical spin chains whose universal behaviors are governed by unitary conformal field theories~(CFTs), the entanglement spectra of subsystems in certain geometries have been shown to be given by a boundary CFT spectrum with appropriately chosen boundary conditions~(bcs)~\cite{Lauchli2013, Ohmori:2014eia, Cardy2016}. For a chain with~$L$ sites and identical bc~($b$) at the ends, the lowest energy of the entanglement spectrum of a subsystem occupying the left~$l$ sites is described by a large-$L$ asymptotic formula keeping~$l/L$ fixed:
\begin{equation}
\varepsilon_0 = \frac{c}{24\pi}W_L + C + o(1).
\end{equation}
Here,~$c$ is the central charge of the corresponding CFT and
\begin{align}
\label{eq:W_L}
W_L = \ln\left(\frac{2L}{\pi}\sin\frac{\pi l}{L}\right).
\end{align}
The constant term~$C$ depends on the details of the lattice model. However, for two different bcs~$b', b$, the corresponding difference~$C' - C$ is universal and expressed in terms of the Affleck-Ludwig g-functions~\cite{Affleck1991}:
\begin{equation}
C'- C = \frac{1}{2\pi}\ln\frac{g_{b'}}{g_{b}}.
\end{equation}
As argued in Ref.~\cite{Cardy2016}, restricting to the class of the low-energy states, the entanglement Hamiltonian possesses the following scaling limit:
\begin{equation}
\label{eq:H_ab}
{\rm s}\!\!\!\lim\limits_{L\rightarrow \infty}W_L\left(K_A - \varepsilon_0\right) = H_{ab},
\end{equation}
where the symbol~${\rm s}\!\lim$ is used to emphasize that the limit is performed for the low-energy states with the size of the system going to infinity keeping the ratio~$l/L$ fixed. Remarkably, the spectrum of~$H_{ab}$ coincides with that of a conformal Hamiltonian with~$a$ and $b$ being obtained from the bcs at the `entanglement cut' and the physical end of the system respectively. Even without specialization of the conformal bc~$a$, the universal part of the spectrum comprises equidistant levels organized into conformal towers. This is reminiscent of the spectrum of Baxter's corner transfer matrices for integrable, non-critical statistical mechanics models~\cite{Baxter2013}.

A natural question is: for what choice of~$a$ would the spectrum of~$H_{ab}$ describe the universal behavior of the entanglement Hamiltonian? A conformal bc is widely accepted to be the `natural choice'~\cite{Cardy2016}. In fact, lattice computations for the entanglement spectra in the Ising, Potts and free boson models have corroborated the conformal nature of the bc~$a$~\cite{Lauchli2013, Ohmori:2014eia, Roy2020a}. For these models,~$a$ has been numerically found to be the free bc. However, it is not immediately obvious how to generalize this result to other interacting CFTs. For unitary CFTs, due to the existence of a monotonic RG flow with decreasing boundary entropy~\cite{Affleck1995conformal, Friedan2004}, one might expect~$a$ to correspond to the bc that has the the lowest boundary entropy and yet does not violate the global symmetries of the bulk model. Indeed, this would explain the numerical results obtained for Ising, Potts and free boson models~\cite{Lauchli2013, Ohmori:2014eia, Roy2020a}. 

\begin{figure*}
\centering
\includegraphics[width = \textwidth]{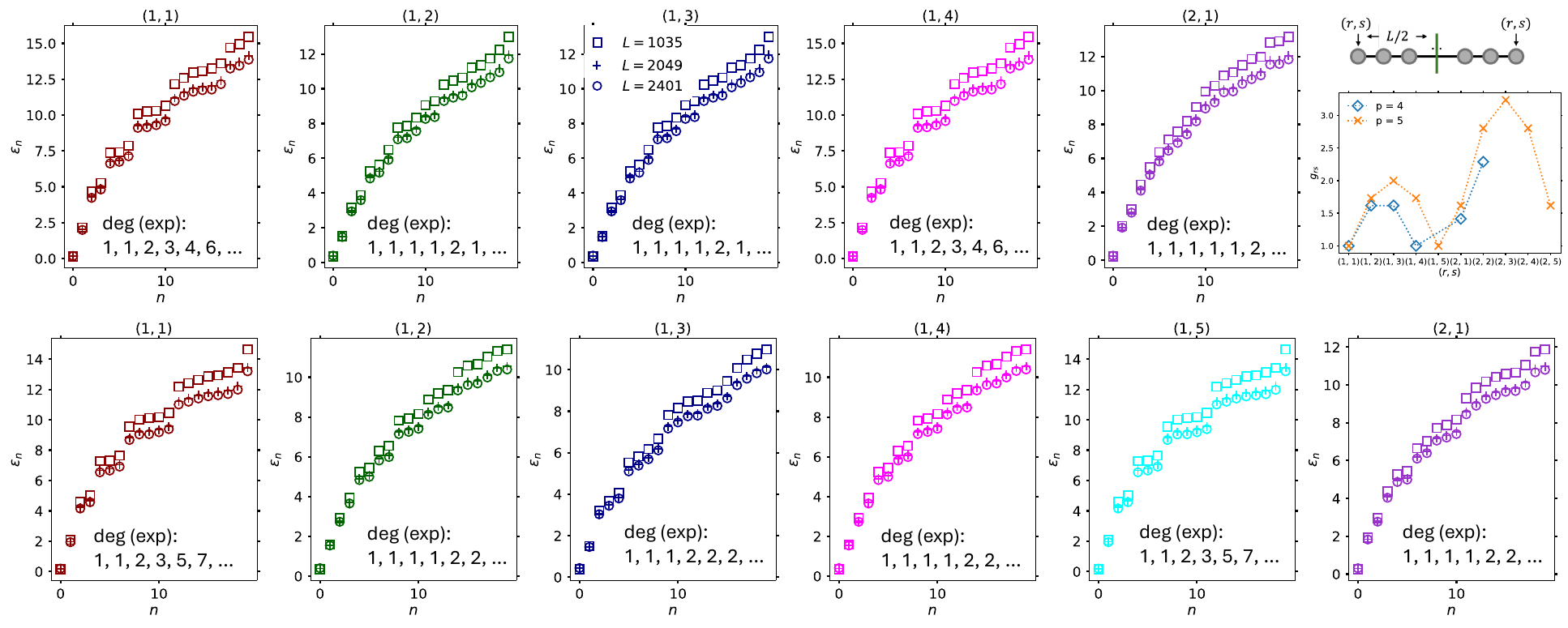}
\caption{\label{fig_1} DMRG results for the entanglement spectra of subsystem of size~$l = L/2$ for the~$A_4$~(upper five panels) and~$A_5$~(lower six panels) RSOS models with $L = 1035, 2049$ and $2401$ sites. The total system is in the ground state of Eq.~\eqref{eq:H_TL}. The different bcs imposed at both ends of the chain are shown in the titles. The g-functions for the different conformal bcs are shown in the top right panel. The maximum of the g-function occurs for~$(r_a, s_a) = (2,2)$ and~$(3,3)$ for the~$A_4$ and $A_5$ models respectively.  Note that the Kac labels~$(2,3)$ and~(3,3) are identified for~$p = 5$. The degeneracies obtained using DMRG are {\it only} compatible with those obtained from the boundary CFT partition functions if the entanglement cut bc~$(r_a, s_a)$ is chosen following Eq.~\eqref{eq:bc}.}
\end{figure*}

The purpose of this work is to show that the above expectation is not vindicated by numerical computations for the unitary minimal models of the~$A_p$ family~\cite{DiFrancesco:1997nk}. For these models, the boundary condition~$a$ which occurs in Eq.~\eqref{eq:H_ab} is identified with the conformal bc with the {\it highest boundary entropy}. Note that this bc preserves the global~$\mathbb{Z}_2$ symmetry of the bulk CFT. The results are obtained using density matrix renormalization group~(DMRG) computations~\cite{Tenpy2024} of ground states of open restricted solid-on-solid~(RSOS) spin chains~\cite{Andrews:1984af, Saleur:1990uz} of the~$A_p$ family. The latter realize the unitary minimal 2D CFTs~${\cal M}(p+1, p)$,~$p = 3,4,5,\ldots$ in the scaling limit. Several of the conformal bcs have been identified at the lattice level for these models~\cite{SALEUR1989591, Cardy1989}. This allows systematic investigation of the different cases and verification of the nature of the bc for the `entanglement cut'. The results obtained for the~$A_4$ RSOS model are found to be compatible with those for the Blume-Capel model~\cite{Blume1966, Capel1966}, providing further evidence regarding the universal nature of the obtained results.

The Hamiltonian for an open RSOS spin chain with~$L$ sites is given by~\cite{Andrews:1984af, Baxter2013, Saleur:1990uz}
\begin{equation}
\label{eq:H_TL}
H = -\frac{\gamma}{\pi\sin\gamma}\sum_{j = 1}^{L - 2}e_j,
\end{equation}
where~$e_j$-s are the generators of the Temperley-Lieb algebra. The allowed states in the Hilbert space for~$L$ sites are of the form~$|a_0, a_1, \ldots, a_{L - 2}, a_{L-1}\rangle$ with $a_i = 1,2,\ldots, p$. For~$A_p$ RSOS models, $a_{i+1}=a_i\pm 1$ and~$\gamma = \pi/(p + 1)$. The DMRG computation is performed after rewriting the Temperley-Lieb generators as projectors~\cite{Roy:2024xdi}:
\begin{align}
\label{eq:TL_op}
e_j = \sum_{a = 1}^p P_{j-1}^{(a)}\tilde{e}^{(a)}_j P_{j +1}^{(a)},
\end{align}
where
\begin{align}
\sqrt{\phi_a}P_j^{(a)} = |a\rangle_j\langle a|,\ \tilde{e}_j^{(a)} = \!\!\!\!\sum_{b_1, b_2=a\pm1}\sqrt{\phi_{b_1}\phi_{b_2}}|b_1\rangle_j\langle b_2|
\end{align}
and~$\phi_a = \sqrt{2\gamma/\pi}\sin(a\gamma)$,~$a = 1, \ldots, p$. The couplings in Eq.~\eqref{eq:H_TL} are chosen such that the Hamiltonian is fine-tuned to the quantum critical point described by the~${\cal M}(p+1,p)$ CFT. For the latter model, the bc~$(1, s)$, $s = 1, \ldots, p$ is realized by fixing the boundary spin in the state~$s$~\cite{SALEUR1989591}. Similarly, the bc~$(r,1)$ is realized by fixing the two spins closest to the boundary in the states~$r, r+1$,~$r = 1, \ldots, p$. The g-function for the bc~$(r,s)$ is given by~\cite{DiFrancesco:1997nk}:
\begin{equation}
\label{eq:g_rs}
g_{rs} = \frac{\sin\frac{\pi r}{p}\sin\frac{\pi s}{p+1}}{\sin\frac{\pi}{p}\sin\frac{\pi}{p+1}}.
\end{equation}
Clearly, the maximum value of the g-function is reached for 
\begin{align}
\label{eq:bc}
r_a = s_a = \begin{cases}
\frac{p}{2},\ &p = {\rm even}\\[0.1cm]
\frac{p+1}{2},\ &p = {\rm odd}
\end{cases}
\end{align}
These bcs have been dubbed `quasi-free'~\cite{Behrend:2000us} and correspond to the most unstable boundary fixed point in the renormalization group sense~(see Ref.~\cite{Cappelli_2004} for details). These are the bcs that describe the universal properties of the entanglement cut in the~$A_p$ models. Notice that the free bc,~$(2,2) = (1,2)$, is recovered for the Ising case~($p=3$). For the tricritical Ising model~($p=4$), Eq.~\eqref{eq:bc} implies the bc~$(2,2)$, for~$p = 5,6$, the relevant bc is~$(3,3)$, and so on~(see top right panel of Fig.~\ref{fig_1}). 

The above claim is validated numerically. The entanglement spectra are obtained from the Schmidt eigenvalues for a given bipartition of the total system, which are readily accessible in a DMRG computation. Even though the physical states of the RSOS models obey a set of constraints, the Schmidt decomposition can be performed without encountering the complications that arise in models with gauge symmetries~\cite{Casini:2013rba}. To verify that this leads to meaningful results, the DMRG simulations were cross-checked for small system-sizes with exact computations. The latter perform the Schmidt decomposition in bases which only contain the states allowed by the RSOS constraints. This is in contrast to the DMRG computations relying on Eq.~\eqref{eq:TL_op} which are defined in the larger tensor-product Hilbert space of~$p$-level spins at each site. The bcs~$(1,s)$ are implemented by choosing the initial state for the DMRG simulations as~$|s, \ldots, s\rangle$. In contrast, for the bc~$(2,1)$, the initial state is chosen to be~$|2,3,\ldots, 3,2\rangle$ and a term~$-h_b|2,3\rangle\langle2,3|$ added to the Hamiltonian on the first two and the last two sites of the chain. The boundary field strength~$h_b$ was taken to be large enough so that the associated length-scale is much smaller than the system-size~\cite{Affleck2009}. The DMRG computation was performed while conserving the total overall parity, where each state~$|a\rangle$ at a given site is assigned a parity~$(-1)^a$. In each case, the occupation probability of a given state~$|a\rangle$ for the~$j^{\rm th}$ site~(expectation value of the projector~$|a\rangle_j\langle a|$,~$a = 1, \ldots, p$) was computed to verify that the correct bc was realized. Furthermore, the ground state energy scaling was verified to agree with the CFT predictions~\cite{SALEUR1989591}. 

A direct comparison between the spectrum of the conformal Hamiltonian with the entanglement energies for finite-size numerical simulations is difficult due to the logarithmic scaling~[Eqs.~(\ref{eq:W_L}, \ref{eq:H_ab})]. However, the degeneracies of the low-lying entanglement spectra are organized into multiplets that correspond to the proposed boundary CFT spectrum. Fig.~\ref{fig_1} presents the entanglement spectra for the left half of an open chain with identical bcs at the ends are shown for the~$A_4$~(upper five panels) and the~$A_5$~(lower six panels) for three different system-sizes~$L = 1035, 2049$ and $2401$. The results are shown for bcs~$(1,s)$, $s = 1, \ldots, p$ and~$(2,1)$. Due to the even-odd parity constraint of the allowed states in the Hilbert space of the~$A_p$ RSOS models, identical bc at the ends are realized only for chains of odd length. The expected degeneracies, straightforwardly obtained from the explicit expressions for the Cardy states~\cite{Cardy1989} and the characters of the Virasoro algebra~\cite{DiFrancesco:1997nk}, are quoted for each case. Importantly, the obtained numerical data is consistent with CFT predictions {\it only} after the bc for the entanglement cut is identified as prescribed in Eq.~\eqref{eq:bc}. Similar results were also obtained for other values of~$p\leq8$. Note that the entanglement spectra for the bcs~$(1, s)$ and~$(1, p + 1-s)$ are identical, as expected from symmetry considerations in the Ginzburg-Landau formulation of these models~\cite{Zamolodchikov:1986db}. 

The identification of the relevant bc for the entanglement cut also enables determination of the exponent governing the `unusual corrections' to entanglement entropies~\cite{Cardy2010}. For the case considered here, the entanglement entropy is given by~$S = cW_L/6 + S_0$, where~$S_0$ is an~${\cal O}(1)$ term that contains non-universal contributions. As the bc at the ends of the chain is varied from~$b$ to~$b'$, the change in~$S_0$ is determined by the corresponding g-functions:~$\Delta S_{\rm CFT} = \ln(g_b/g_b')$. For large~$W_L$, the leading order correction to the last equation is given by~${\cal O}(e^{-2\nu W_L})$, where~$\nu$ is determined by the operator spectrum of the CFT with bcs~$(r,s)$ and~$(r_a, s_a)$~\cite{Cardy2010, Cardy2016}. For~$l = L/2$, these corrections decay as~$~L^{-2\nu}$~[Eq.~\eqref{eq:W_L}]. 

\begin{figure}
\centering
\includegraphics[width = 0.45\textwidth]{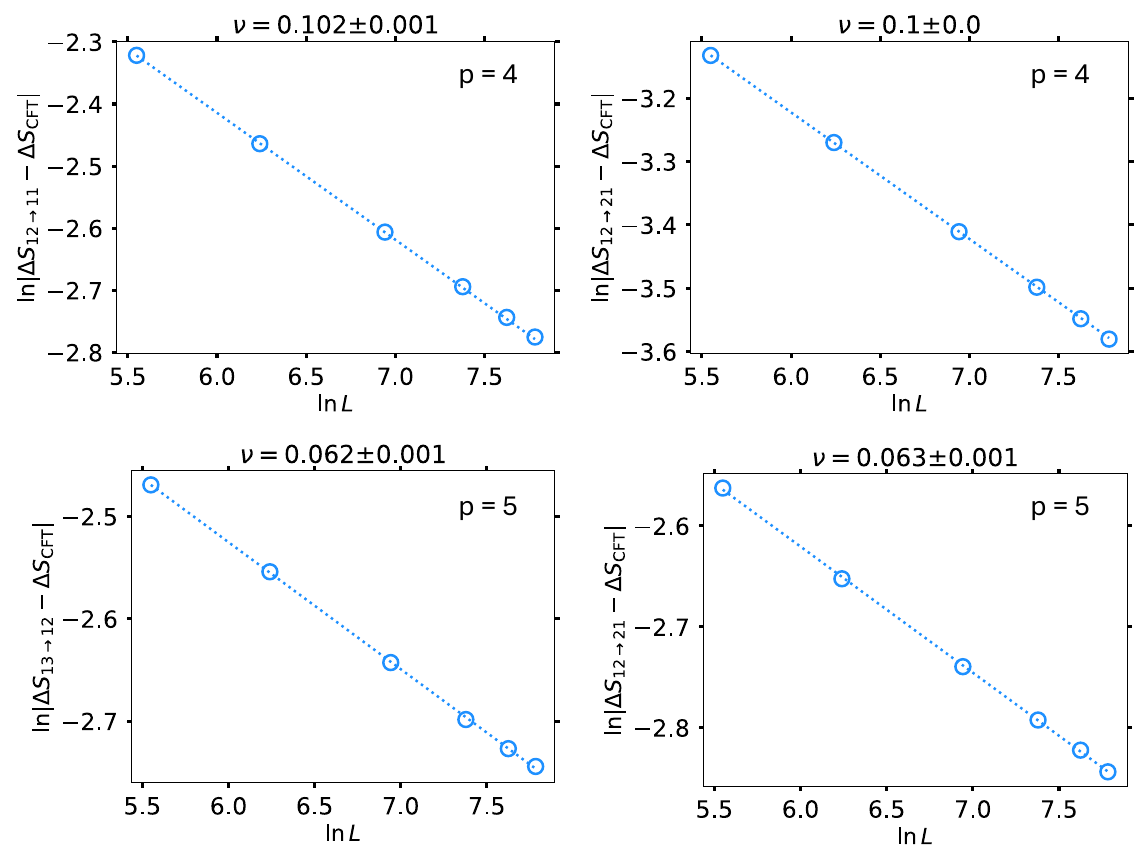}
\caption{\label{fig_2} Results for the unusual corrections to half-chain entanglement entropy for the~$A_4$~(upper panels) and~$A_5$~(lower panels) RSOS models. The change~$\Delta S$ is computed using DMRG by changing the  imposed bc at the ends of the system~(indicated by the subscript in the y-labels).  The exponents of the unusual corrections to the entanglement entropy are obtained by fitting to Eq.~\eqref{eq:S_un}. In the latter,~$\Delta S_{\rm CFT}$ is calculated for the different bcs using Eq.~\eqref{eq:g_rs}. For~$p = 4(5)$, the obtained exponent~$\nu$ is close to the expected result of~$1/10(1/15)$ arising from the~$(3,3)$ field of the corresponding CFT. The expected results were determined from the corresponding partition functions.}
\end{figure}

Next, the exponent~$\nu$ is obtained by numerically computing~$\Delta S$ and using
\begin{equation}
\label{eq:S_un}
|\Delta S - \Delta S_{\rm CFT}|\sim L^{-2\nu}.
\end{equation}
Fig.~\ref{fig_2} shows the DMRG results for the~$A_4$ and~$A_5$ models. The difference in the entanglement entropy of the left half of the chain for different system-sizes as the bcs at the ends are changed is first computed using DMRG. The corresponding CFT predictions are obtained from ratio of the relevant g-functions~[Eq.~\eqref{eq:g_rs}]. Fitting the deviation to Eq.~\eqref{eq:S_un} yields the exponent~$\nu\simeq 0.1~(0.063)$ for~$A_4(A_5)$ models. This is close to what is expected from the boundary CFT partition functions, which predict~$\nu = 1/10(1/15)$ arising from the~$(3,3)$ field in the corresponding CFT. These exponents are also compatible with the corner transfer matrix results for the off-critical case done in Ref.~\cite{DeLuca2013}. 

Finally, the bc for the entanglement cut is investigated using a different lattice realization for the minimal model~${\cal M}(5,4)$, namely the Blume-Capel spin chain~\cite{Blume1966, Capel1966} for a specific choice of parameters, see Eqs.~(\ref{eq:H_BC}, \ref{eq:BC_p}) below. This is done to provide additional numerical evidence to support the claim that the results obtained in this work do correspond to the underlying CFT and not a consequence of specific lattice realizations. This is important since the entanglement spectra is known to sometimes exhibit non-universal characteristics for models in the same phase~\cite{Chandran2014}. However, as shown below, the bc that best captures the universal properties of the entanglement cut is still~$(2,2)$. 

The lattice Hamiltonian for the Blume-Capel model is given by~\cite{vonGehlen1990}:
\begin{align}
\label{eq:H_BC}
H_{\rm BC} &= \xi\sum_{j = 1}^{L}\left[\alpha(S_j^x)^2 + \beta S^z_j + \gamma (S_j^z)^2\right] - \xi\sum_{j = 1}^{L - 1}S_j^xS_{j+1}^x\nonumber\\&\qquad - h_bS_1^x - h_bS_{L}^x,
\end{align}
where~$S_j^{x,z}$ are spin-1 matrices. For the tricritical point, the parameters~$\alpha,\beta,\gamma$ and~$\xi$ are chosen to be~\cite{vonGehlen1990}:
\begin{equation}
\label{eq:BC_p}
\alpha = 0.910207, \beta = 0.415685, \gamma = 0, \xi = 1/0.56557.
\end{equation}
The overall scale-factor~$\xi$ is chosen such that the energy spectra is normalized to match the CFT results. The~$(2,1)$ bc at the ends is obtained by choosing~$h_b = 0$~\cite{Balbao1987, Chim:1995kf, Affleck:2000jv}. The~$(1,1)$ bc is realized when~$h_b$ is chosen to be a positive value that is large enough so that the associated correlation length is much smaller than the system-size. The~(1,4) bc is obtained by changing the sign of~$h_b$ used in the last case. 

\begin{figure}
\centering
\includegraphics[width = 0.49\textwidth]{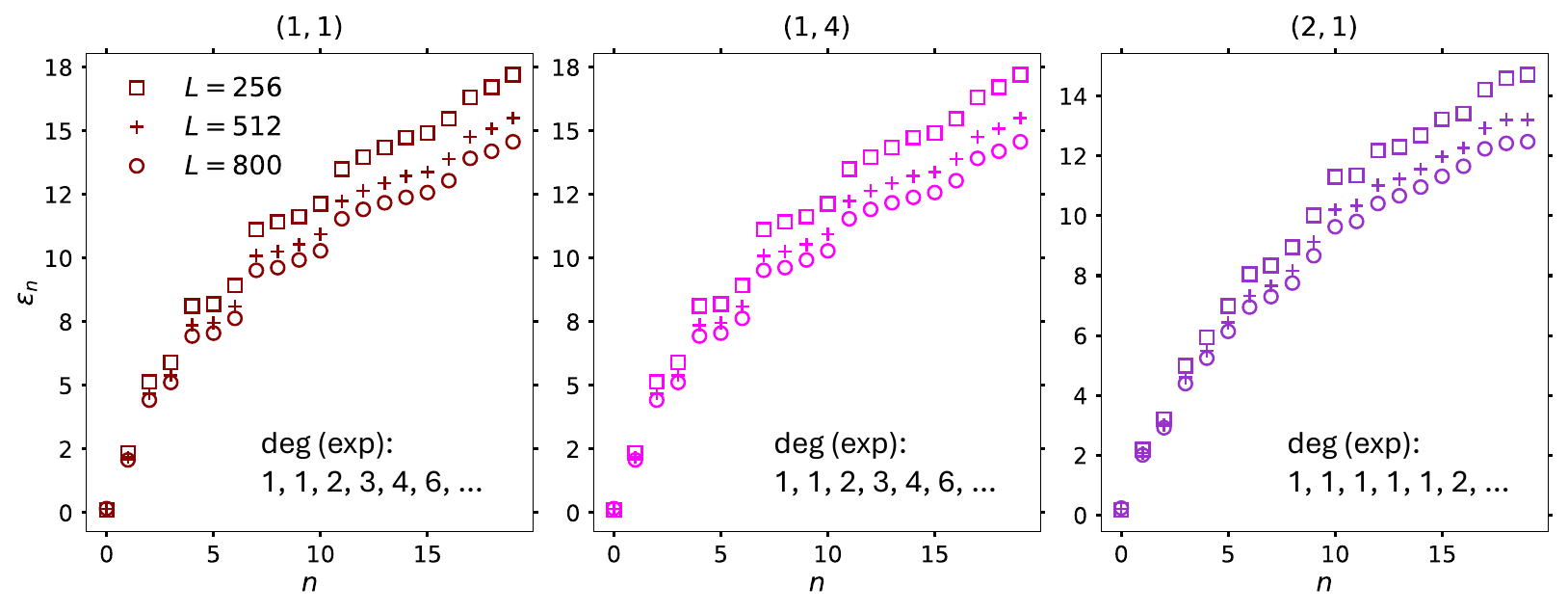}
\caption{\label{fig_3} DMRG results for the entanglement spectra of a subsystem of size~$l = L/2$ for the Blume-Capel model with~$L = 256, 512$ and~$800$ sites~(compare Fig.~\ref{fig_1}, upper panels). The bcs imposed~(identical at both ends) are indicated in the titles. The expected degeneracies are also shown. As for the~$A_4$ RSOS model, the CFT predictions agree with the DMRG results {\it only} if the bc for the entanglement cut is identified with~$(2,2)$.}
\end{figure}

Fig.~\ref{fig_3} shows the results for the entanglement spectra for the ground state of the Blume-Capel Hamiltonian~[Eq.~\eqref{eq:H_BC}] with three different bcs. The subsystem size is~$l = L/2$ and the system-size is chosen to be~$L = 256, 512$ and~$800$. The corresponding CFT predictions are also shown. The DMRG results for the low-lying entanglement energy degeneracies are in reasonable agreement with the CFT predictions as long as the bc for the entanglement cut is chosen to be~$(2,2)$. The~$(1,2)$ and~$(1,3)$ bcs were also checked to yield results compatible with those obtained from RSOS realizations. The exponents for the unusual corrections to the entanglement entropy were also found to agree with that obtained for the~$A_4$ RSOS model.

To summarize, the nature of the bc for the entanglement cut was analyzed for ground states of unitary 2D CFTs using the DMRG technique. To that end, the degeneracy of the entanglement spectra was used as a robust diagnostic. For~$A_p$ RSOS models, which realize unitary minimal CFTs in the scaling limit, the bc for the entanglement cut was found to be conformal and coinciding with the one with the highest boundary entropy. This is compatible with the heuristic picture that the entanglement cut should correspond to the bc with the highest number of boundary degrees of freedom. In the Landau-Ginzburg description of conformal bcs, this corresponds as well to the situation with no boundary potential \cite{Cappelli_2004}. The three-state Potts model, analyzed in Ref.~\cite{Lauchli2013}, proves to be an exception to this rule, with the bc for the entanglement cut corresponding to free bc. The latter's g-function is less than that of the `new' bc identified in Ref.~\cite{Affleck:1998nq}. It is possible that the entanglement cut for the~$A_p$ RSOS models simply chooses the bc that is closest to a `free' bc and ends up corresponding to the one with the highest boundary entropy. It is intriguing that these bcs are reached in an entanglement cut even though they are the most unstable in the renormalization group sense and are difficult to realize on the lattice. 

The determination of the nature and explicit form of the bc at the entanglement cut for the different minimal models is essential for quantitative understanding of several key questions concerning entanglement properties of CFTs. These include symmetry-resolution of entanglement entropies for 2D CFTs~\cite{Goldstein2018, Calabrese:2021wvi, Capizzi2020symmetry} and computations of interface entropies in CFTs with topological defect lines~\cite{Roy2021a, Roy2024}. Furthermore, Bethe ansatz computations of the CFT partition functions in the presence of boundaries could be an alternative method for computation of entanglement spectra of those CFTs which are difficult to investigate using other numerical methods. Finally, note that the DMRG computations performed in this work conserve the overall parity of the RSOS Hamiltonians. Conservation of RSOS charges as described in Refs.~\cite{Sierra:1996qz, Pfeifer2015} could enhance the precision of the obtained results and enable sorting of the entanglement spectra into different conformal towers. 

Discussions with Pasquale Calabrese and Erik Tonni are gratefully acknowledged. The work of H.S. was supported by the French Agence Nationale de la Recherche (ANR) under grant ANR-21- CE40-0003 (project CONFICA). S.L. was supported by NSF-PHY-2210187. 

\bibliography{/Users/ananda/Dropbox/Bibliography/library_1}

\end{document}